\documentclass[useAMS,usenatbib]{mn2e}
\usepackage{multirow}
\usepackage{epsfig}
\usepackage{subfigure}

\makeatletter
\let\jnl@style=\rm
\def\ref@jnl#1{{\jnl@style#1}}

\def\aj{\ref@jnl{AJ}}                   % Astronomical Journal
\def\araa{\ref@jnl{ARA\&A}}             % Annual Review of Astron and Astrophys
\def\apj{\ref@jnl{ApJ}}                 % Astrophysical Journal
\def\apjl{\ref@jnl{ApJ}}                % Astrophysical Journal, Letters
\def\apjs{\ref@jnl{ApJS}}               % Astrophysical Journal, Supplement
\def\ao{\ref@jnl{Appl.~Opt.}}           % Applied Optics
\def\apss{\ref@jnl{Ap\&SS}}             % Astrophysics and Space Science
\def\aap{\ref@jnl{A\&A}}                % Astronomy and Astrophysics
\def\aapr{\ref@jnl{A\&A~Rev.}}          % Astronomy and Astrophysics Reviews
\def\aaps{\ref@jnl{A\&AS}}              % Astronomy and Astrophysics, Supplement
\def\azh{\ref@jnl{AZh}}                 % Astronomicheskii Zhurnal
\def\baas{\ref@jnl{BAAS}}               % Bulletin of the AAS
\def\jrasc{\ref@jnl{JRASC}}             % Journal of the RAS of Canada
\def\memras{\ref@jnl{MmRAS}}            % Memoirs of the RAS
\def\mnras{\ref@jnl{MNRAS}}             % Monthly Notices of the RAS
\def\pra{\ref@jnl{Phys.~Rev.~A}}        % Physical Review A: General Physics
\def\prb{\ref@jnl{Phys.~Rev.~B}}        % Physical Review B: Solid State
\def\prc{\ref@jnl{Phys.~Rev.~C}}        % Physical Review C
\def\prd{\ref@jnl{Phys.~Rev.~D}}        % Physical Review D
\def\pre{\ref@jnl{Phys.~Rev.~E}}        % Physical Review E
\def\prl{\ref@jnl{Phys.~Rev.~Lett.}}    % Physical Review Letters
\def\pasp{\ref@jnl{PASP}}               % Publications of the ASP
\def\pasj{\ref@jnl{PASJ}}               % Publications of the ASJ
\def\qjras{\ref@jnl{QJRAS}}             % Quarterly Journal of the RAS
\def\skytel{\ref@jnl{S\&T}}             % Sky and Telescope
\def\solphys{\ref@jnl{Sol.~Phys.}}      % Solar Physics
\def\sovast{\ref@jnl{Soviet~Ast.}}      % Soviet Astronomy
\def\ssr{\ref@jnl{Space~Sci.~Rev.}}     % Space Science Reviews
\def\zap{\ref@jnl{ZAp}}                 % Zeitschrift fuer Astrophysik
\def\nat{\ref@jnl{Nature}}              % Nature
\def\iaucirc{\ref@jnl{IAU~Circ.}}       % IAU Cirulars
\def\aplett{\ref@jnl{Astrophys.~Lett.}} % Astrophysics Letters
\def\apspr{\ref@jnl{Astrophys.~Space~Phys.~Res.}}
\def\bain{\ref@jnl{Bull.~Astron.~Inst.~Netherlands}}
\def\fcp{\ref@jnl{Fund.~Cosmic~Phys.}}  % Fundamental Cosmic Physics
\def\gca{\ref@jnl{Geochim.~Cosmochim.~Acta}}   % Geochimica Cosmochimica Acta
\def\grl{\ref@jnl{Geophys.~Res.~Lett.}} % Geophysics Research Letters
\def\jcp{\ref@jnl{J.~Chem.~Phys.}}      % Journal of Chemical Physics
\def\jgr{\ref@jnl{J.~Geophys.~Res.}}    % Journal of Geophysics Research
\def\jqsrt{\ref@jnl{J.~Quant.~Spec.~Radiat.~Transf.}}
\def\memsai{\ref@jnl{Mem.~Soc.~Astron.~Italiana}}
\def\nphysa{\ref@jnl{Nucl.~Phys.~A}}   % Nuclear Physics A
\def\physrep{\ref@jnl{Phys.~Rep.}}   % Physics Reports
\def\physscr{\ref@jnl{Phys.~Scr}}   % Physica Scripta
\def\planss{\ref@jnl{Planet.~Space~Sci.}}   % Planetary Space Science
\def\procspie{\ref@jnl{Proc.~SPIE}}   % Proceedings of the SPIE

\makeatother

\title[\textit{Chandra} unveils a binary Active Galactic Nucleus in Mrk~463]{\textit{Chandra} unveils a binary Active Galactic Nucleus in Mrk~463}

\author[Stefano Bianchi, et al.]{Stefano Bianchi$^1$\thanks{E-mail: bianchi@fis.uniroma3.it (SB)}, Marco Chiaberge$^2$, Enrico Piconcelli$^3$, Matteo Guainazzi$^4$,
\newauthor
Giorgio Matt$^1$\\
$^1$Dipartimento di Fisica, Universit\`a degli Studi Roma Tre, via della Vasca Navale 84, 00146 Roma, Italy\\
$^2$Space Telescope Science Institute, 3700 San Martin Drive, Baltimore, MD 21218\\
$^3$Osservatorio Astronomico di Roma (INAF), Via Frascati 33, I-00040 Monte Porzio Catone, Italy\\
$^4$XMM-Newton Science Operations Center, European Space Astronomy Center, ESA, Apartado 50727, E-28080 Madrid, Spain\\
}

\begin{document}

%\pagerange{\pageref{firstpage}--\pageref{lastpage}} \pubyear{2004}

\maketitle

\label{firstpage}

\begin{abstract}
We analyse \textit{Chandra}, XMM-\textit{Newton} and \textit{HST} data of the double-nucleus Ultraluminous Infrared Galaxy (ULIRG), Mrk~463. The \textit{Chandra} detection of two luminous ($\mathrm{L}_\mathrm{2-10\,keV}=1.5\times10^{43}$ and $3.8\times10^{42}$ erg cm$^{-2}$ s$^{-1}$), unresolved nuclei in Mrk~463 indicates that this galaxy hosts a binary AGN, with a projected separation of $\simeq3.8$ kpc ($3.83\pm0.01$ arcsec). While the East nucleus was already known to be a Seyfert 2 (and this is further confirmed by our \textit{Chandra} detection of a neutral iron line), this is the first unambiguous evidence in favour of the AGN nature of the West nucleus. Mrk~463 is therefore the clearest case so far for a binary AGN, after NGC~6240.

\end{abstract}

\begin{keywords}
galaxies: active - galaxies: Seyfert - X-rays: individual: Mrk463 - X-rays: individual: Mrk463E - X-rays: individual: Mrk463W
\end{keywords}

\section{Introduction}

If, as commonly believed, galaxies merge hierarchically and all galactic bulges contain supermassive
Black Holes (BHs), the formation of binary/multiple BHs should be
therefore inevitable \citep[e.g.][]{mm01,hq04}. Interestingly, the presence of binary supermassive BHs
has been also invoked to account for many important aspects of the AGN phenomenon \citep[see][for a review]{kom03}, i.e. the formation of the molecular torus (key ingredient of the Unified models), the difference between radio-loud and radio-quiet AGN, the distortions and the bendings in radio jets, and the random orientations of the radio jets and bi-conical narrow-line regions with respect to the
rotational axes of the host galaxy disk. Furthermore, coalescing binary BHs are expected to be the most powerful sources of gravitational waves.

However, observational evidence for binary BHs is so far very rare. The first clear-cut example was found in the Ultraluminous Infrared Galaxy \citep[ULIRG:][]{sm96}, NGC~6240 \citep{kom03b}. The active nature of the two nuclei is unambiguous and their projected distance ($\simeq1$ kpc) remains the shortest measured so far for such a system. A second, close binary AGN ($\simeq4.6$ kpc) was later claimed in Arp~299 \citep{ballo04}. A system with a larger projected distance between the nuclei ($\simeq10.5$ kpc) was then found in the galactic pair ESO~509-IG066 \citep{gua05c}. Finally, \citet{evans08} revealed the AGN nature of the companion of the FRII radio source 3C~321. 

Mrk~463 (z=0.0504) is an ULIRG with a double nucleus, Mrk~463W and Mrk~463E, and tidal tails due to a recent merger between two spiral galaxies \citep[][and references intherein]{mazz91}. The Eastern nucleus is commonly classified as a Seyfert 2 \citep[e.g.][]{so81,hn89}, with broad optical lines detected in polarized light \citep{mg90}. On the other hand, the nature of the Western nucleus is ambiguous: a Seyfert 2, a LINER or a powerful starbust galaxy are all possible \citep[e.g.][]{so81,mazz91}. In the X-rays, Mrk~463 was not detected by \textit{Ginga} \citep[$\mathrm{L}_\mathrm{2-10\,keV}<2\times10^{43}$ erg s$^{-1}$:][]{awaki93}, but it was detected in the soft X-rays by \textit{Einstein} and \textit{ROSAT} \citep{polletta96}. \textit{ASCA} revealed a very absorbed spectrum \citep{ueno96}, confirmed by \textit{BeppoSAX}, together with the detection of a strong iron line \citep{lb01}. These results were further refined by the XMM-\textit{Newton} observation \citep{it04}.

In this paper, we present a \textit{Chandra} observation of Mrk~463, where the two nuclei are spatially resolved for the first time in the X-rays, performing a detailed comparison with \textit{HST} optical and near infrared (NIR) data. 

\section{Observations and data reduction}

In the following, errors correspond to the 90\% confidence level for one interesting parameter ($\Delta \chi^2 =2.71$), where not otherwise stated. The adopted cosmological parameters are $H_0=70$ km s$^{-1}$ Mpc$^{-1}$, $\Omega_\Lambda=0.73$ and $\Omega_m=0.27$ \citep[i.e. the default ones in \textsc{xspec 12.3.1}:][]{xspec}. At the distance of Mrk~463, 1 arcsec corresponds to 986 pc. In all the fits, the Galactic column density along the line of sight to Mrk~463 is included \citep[$2.06\times10^{20}$ cm$^{-2}$: ][]{dl90}.

\subsection{X-rays: \textit{Chandra} and \textit{XMM-Newton}}

Mrk~463 was observed by \textit{Chandra} on 2004, June 11 (obsid 4913), with the Advanced CCD Imaging Spectrometer \citep[ACIS:][]{acis}.
 Data were reduced with the Chandra Interactive Analysis of Observations \citep[CIAO:][]{ciao} 3.4 and the Chandra Calibration Data Base (CALDB) 3.4.1 software, adopting standard procedures, for a final net exposure time of about 49 ks. Images were corrected for known aspect offsets, reaching a nominal astrometric accuracy of 0.6 arcsec (at the 90 per cent confidence level). Three extraction regions were used for the \textit{Chandra} data: two circular regions with radius of 2 arcsec centered at the hard X-ray nuclei and a circular region with radius of 7 arcsec, encompassing all the soft X-ray extended emission, but excluding the two previous regions. Spectra were re-binned to have at least 25 counts in each bin, in order to use the $\chi^2$ statistics. Local fits were also performed in the 5.5-7.0 keV energy range with the unbinned spectra, using the \citet{cash76} statistics, in order to better assess the presence of the iron lines and measure their properties. We refer the reader to \citet{gua05b} for details on this kind of analysis.

Mrk~463 was observed by XMM-\textit{Newton} on 2001, December 22 (obsid 0094401201), with the EPIC CCD cameras, the pn \citep{struder01} and two MOS \citep{turner01}, operated in Full Frame and Medium Filter. These data were already presented by \citet{it04}, but we reduced them with the latest software and calibration files in order to do a better comparison with the \textit{Chandra} spectrum. Source extraction radii and screening for intervals of flaring particle background were performed with SAS 7.1.0 \citep{sas610} via an iterative process which leads to a maximization of the Signal-to-Noise Ratio (SNR), similarly to what described in \citet{pico04}. After this process, the net exposure time was of about 21, 25 and 26 ks for pn, MOS1 and MOS2 respectively, adopting extraction radii of 28 arcsec for all the cameras. The background spectra were extracted from source-free circular regions with a radius of 50 arcsec. Pattern 0 to 4 were used for the pn spectrum, while MOS spectra include patterns 0 to 12. Spectra were binned in order to oversample the instrumental resolution by at least a factor of 3 and to have no less than 30 counts in each background-subtracted spectral channel.

\subsection{Optical and NIR: \textit{HST}}

\textit{HST} observations were retrieved from the Multimission Archive at STScI
and  processed through  the standard  on-the-fly  reprocessing system.
Mrk~463 was observed  with WFPC2, NICMOS and with  various optical and
NIR filters, as part  of different observational programs.  We use
NIR observations with NICMOS-NIC2 (NIR filters F110W, F160W, F207M)
from program GO~7213. We also use optical observations with WFPC2 from
program GO~5982  (F814W filter),  GO~6301 (F588N and  FR533N filters).
The images from GO~6301 are  combined to remove cosmic rays using the
task {\it  crrej} in  {\it IRAF}, and  bad columns were removed using
{\it fixpix}.  For the F814W  filter (left panel of fig. \ref{nuclei}), only one exposure was
taken. Therefore, in order to remove cosmic rays and hot pixels, we run
an iterative process using the tasks {\it cosmicrays} and {\it fixpix}
to identify cosmic rays and hot pixels and then `fix' the image.

The  image  taken with  the  ramp filter  FR533N  is  centered at  the
redshifted wavelength  of the [O\,\textsc{iii}]$\lambda5007\mathrm{\AA}$ emission line,  while the
F588N filter  only includes the nearby continuum  emission.  The image
with F588N is  then rescaled to match the  continuum emission detected
with the  FR533N filter, and  the continuum emission is  subtracted in
order to obtain a `pure' [O\,\textsc{iii}] image (Fig. \ref{oiii2xray}).

We measure  the NIR flux of  the nuclei from the  NICMOS images by
performing  aperture photometry  using  the task  {\it radprof}.   The
aperture  radius  is  set at  $\sim  6$  pixels  from the  PSF  center
(corresponding  to  $\sim  0.45\arcsec$)  for  the  brightest  sources
(e.g. E nucleus in F207M), and  closer to the center ($\sim 4$ pixels)
for  the faintest  (e.g.  W  nucleus  in F110W).   The background  was
measured in an  annulus of 1 pixel width, as close  as possible to the
nucleus, just outside the  aperture radius. Errors on the measurements
depend on the  contrast between the brightness of  the nucleus and the
surrounding background.  We  estimate that errors are of  the order of
5\% (corresponding to the NICMOS photometric calibration error) for the
brightest  sources and  up to  40\%  for the  W nucleus  in F110W.   We
convert  the counts measured inside the aperture into  total  counts  of the  PSF  using
aperture corrections derived by performing similar aperture photometry
on  appropriate  synthetic  PSFs   obtained  with  the  {\it  TinyTim}
software. The counts are then  converted into physical units using the
PHOTFLAM keyword in  the image header as a  first order approximation.
The  PHOTFLAM keyword  assumes a  source with  a flat  continuum slope
($F_\lambda \propto \lambda^\beta$, with $\beta=0$). Thus, in presence
of steep  spectral slopes, the  conversion may be incorrect.  By using
the  task {\it  bandpar} in  {\it SYNPHOT}  we iteratively  derive the
slope  of   the  nuclei,  and   recalculate  the  value   of  PHOTFLAM
corresponding to the measured slope,

The NIR  filters on \textit{HST}/NICMOS  are similar but not  equivalent to
the standard J, H and K filters used in ground based telescopes \citep{barker07}.  In order to compare our measurements with K-band magnitudes published in the literature,
we measure the flux of the nuclei in the NICMOS F207M image, inside an aperture of $1\arcsec$ radius (corresponding to 13.25 pixels). In this case the background is measured far away from the nuclei, at a distance of $\sim 6\arcsec$.
We then convert the fluxes measured from the \textit{HST} image to K-band magnitudes using the task
{\it calcphot} in SYNPHOT. The   observations  were  performed  before  the
cryo-cooler installation  on NICMOS,  therefore we modify  the default
settings  of the task  to make  use of  the correct  system throughput
tables for the time of the observations.

\section{Imaging analysis}

\begin{figure*}
\begin{center}
\epsfig{file=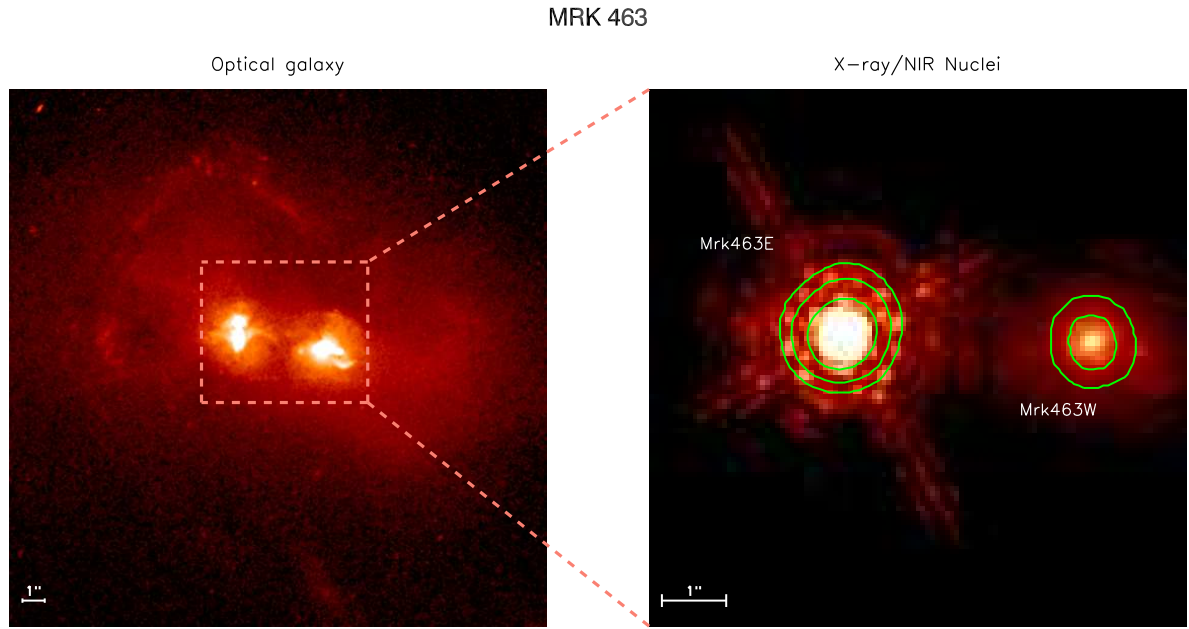, width=17cm}
\end{center}
\caption{\label{nuclei}Mrk~463 (north is up, east to the left). \textit{Left panel}: \textit{HST} optical (filter f814w) image of this double-nucleus galaxy. \textit{Right panel}: hard X-rays \textit{Chandra} contours superimposed on the 2.1$\mu$m \textit{HST} image: two unresolved nuclei are clearly detected in both bands.}
\end{figure*}

\begin{figure}
\begin{center}
\epsfig{file=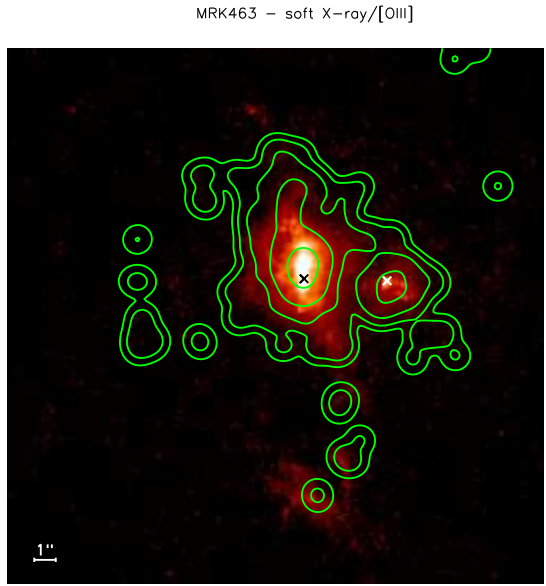, width=8cm}
\end{center}
\caption{\label{oiii2xray}Mrk~463: soft X-rays \textit{Chandra} contours superimposed on the \textit{HST} [{O\,\textsc{iii}}] emission. The two crosses show the positions of the nuclei (see Fig. \ref{nuclei}). North is up, east to the left.}
\end{figure}

The high energy ($>2$ keV) \textit{Chandra} image of Mrk~463 clearly shows the presence of two bright, unresolved nuclei. Their positions are coincident with the two nuclei detected in the \textit{HST} NIR image (Fig. \ref{nuclei}): Mrk~463E (K$=10.98\pm0.05$ mag) and Mrk~463W (K$=13.63\pm0.06$ mag), separated by $3.83\pm0.01$ arcsec. The soft X-ray ($<2$ keV) \textit{Chandra} image still presents two nuclei and some extended emission, which seems more clearly related to the brightest Mrk~463E. The shape of the soft X-ray emission closely resembles the Narrow Line Region (NLR), mapped by the [{O\,\textsc{iii}}] emission (Fig. \ref{oiii2xray}). This is a very common property of Seyfert 2 galaxies and suggests a common origin in a gas photoionized by the AGN emission \citep[see e.g.][]{bianchi06}.

\section{Spectral analysis}

\begin{figure*}
\begin{center}
\epsfig{file=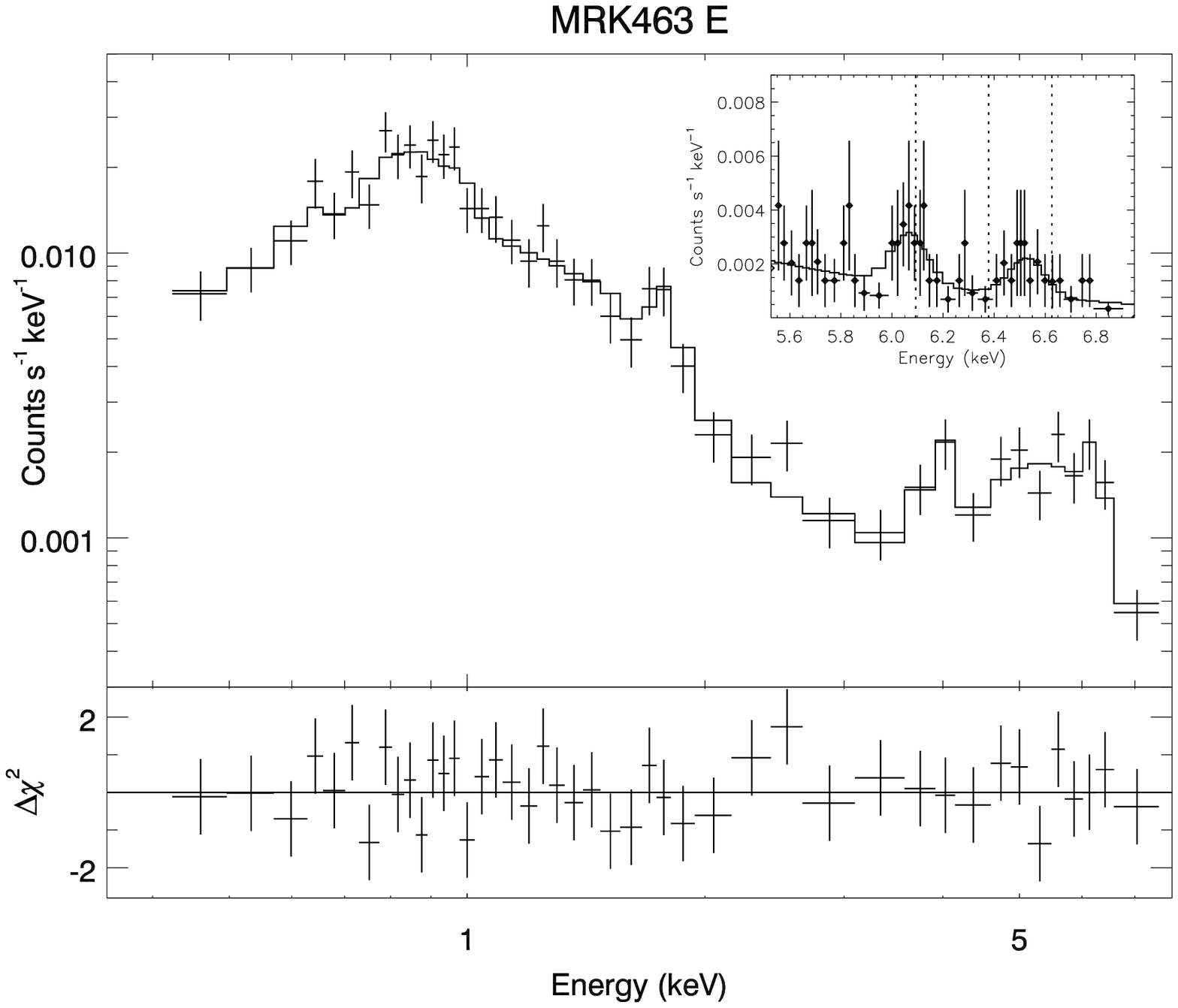, width=8cm}
\epsfig{file=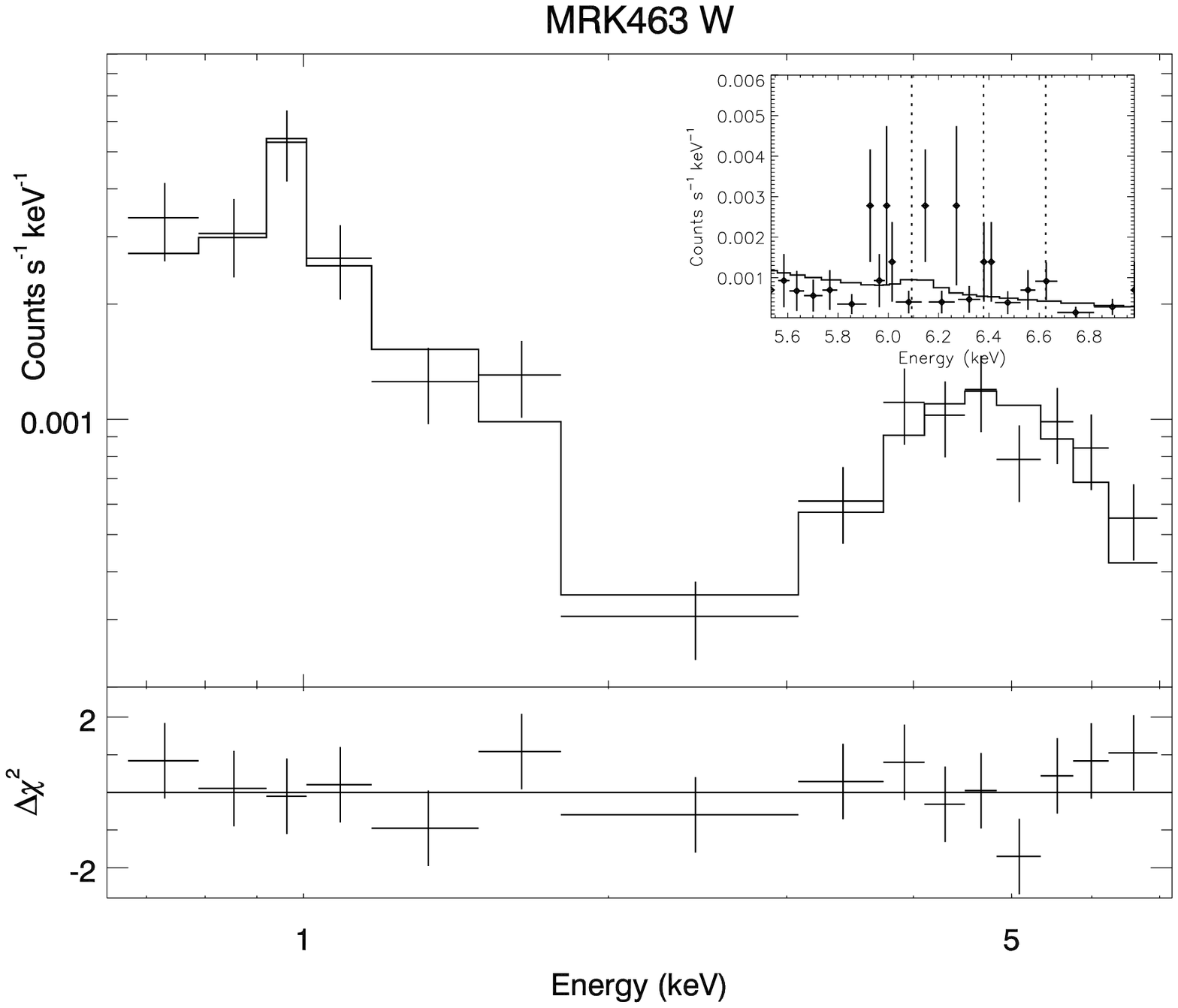, width=8cm}
\epsfig{file=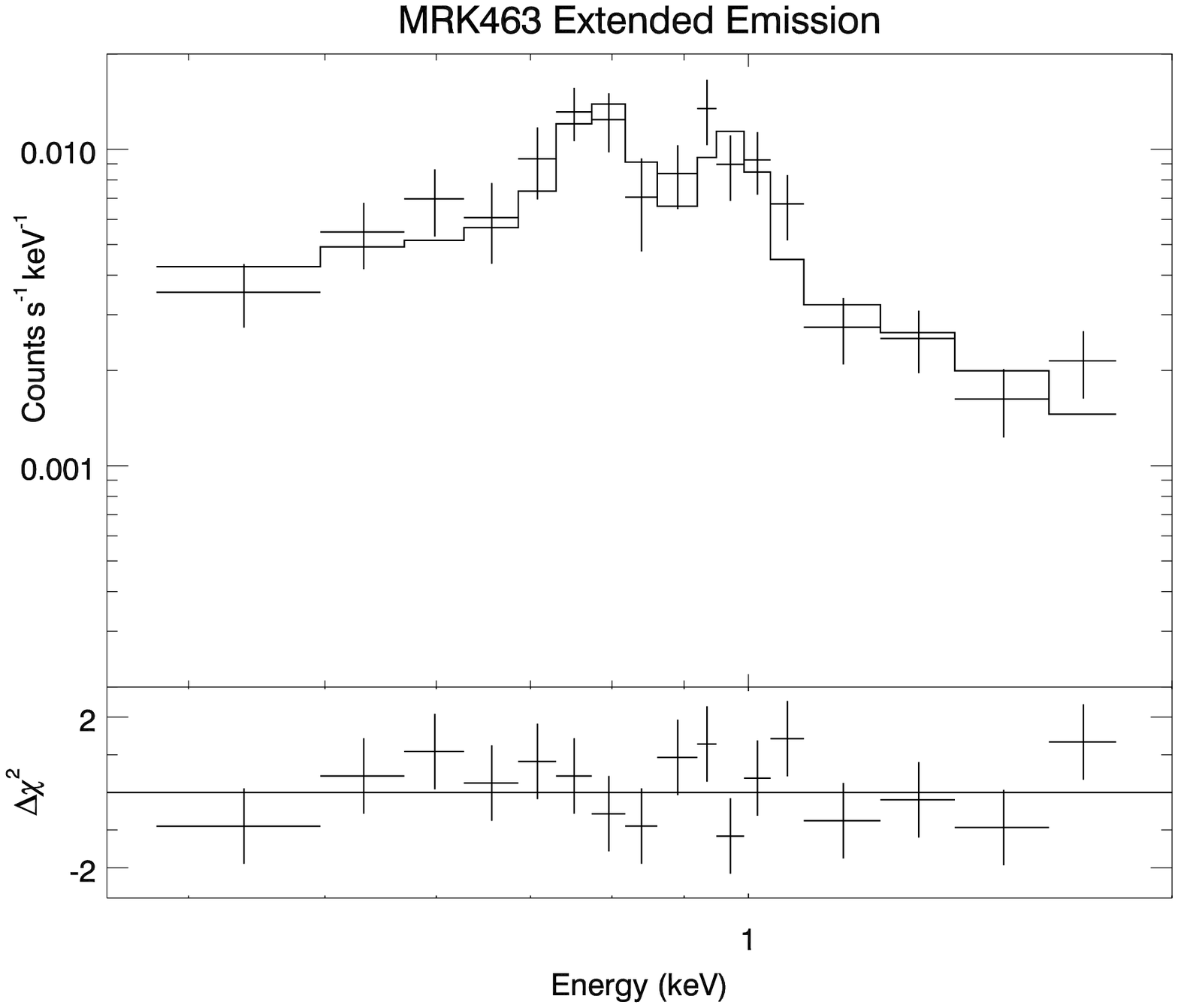, width=8cm}
\epsfig{file=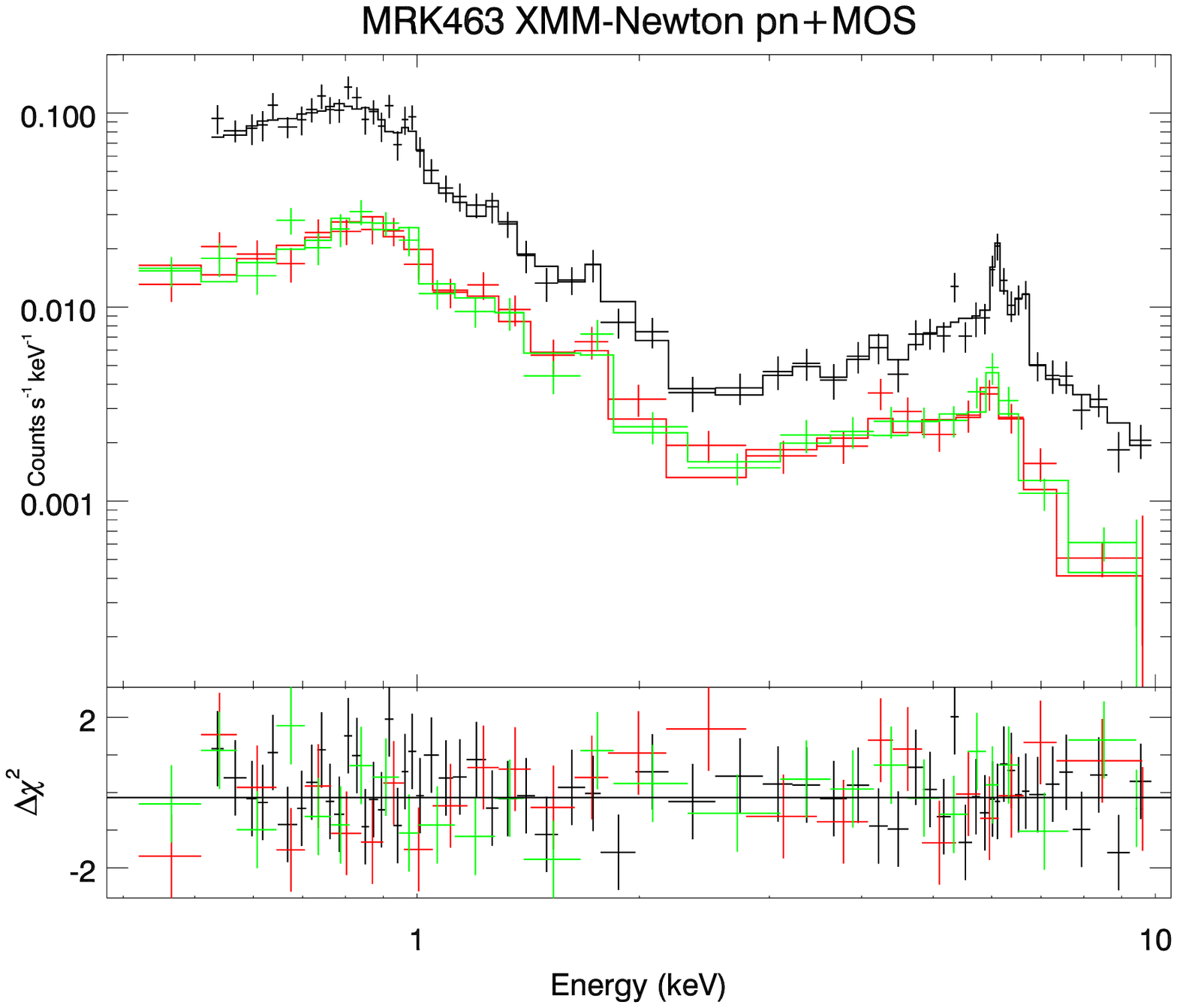, width=8cm}
\end{center}
\caption{\label{chandrafit}\textit{Chandra} spectra of Mrk~463E, Mrk~463W and the extended emission. See text for details. XMM-\textit{Newton} EPIC, MOS1 and MOS2 spectra of Mrk~463.}
\end{figure*}

The \textit{Chandra} spectra of the two nuclei are typical of Compton-thin Seyfert 2 galaxies: they are strongly absorbed, with a soft excess component in the low energy band (see upper panels of Fig. \ref{chandrafit}). They are well fitted by two powerlaws with the same photon index, one of which absorbed by a column density of a few $10^{23}$ cm$^{-2}$. A number of strong emission lines are also required for the East spectrum, while only emission from {Ne\,\textsc{x}} K$\alpha$ is detected in the West spectrum, which has a poorer SNR (see Table \ref{xmmfit}). Interestingly, the soft excess in source E is further absorbed by a local column density of the order of $10^{21}$ cm$^{-2}$, even if with large uncertainty. The spectrum of the extended emission (which excludes the nuclei) is fully consistent with the soft X-ray emission from the nuclei, being well fitted by a simple powerlaw and a couple of emission lines (see lower left panel of Fig. \ref{chandrafit} and Table \ref{xmmfit}). 

\begin{table*}
\caption{\label{xmmfit}Best fit parameters for the \textit{Chandra} and XMM-\textit{Newton} spectra analysed in this paper. Fluxes are in units of $10^{-13}$ erg cm$^{-2}$ s$^{-1}$, luminosities in units of $10^{42}$ erg s$^{-1}$, emission line fluxes in units of $10^{-6}$ ph cm$^{-2}$ s$^{-1}$. In the fits, only significant (at the 90\% level) emission lines were added and, once identified, their centroid energies were fixed to the theoretical values. For the He-like triplets, we fixed an average centroid energy between the three components (r), (i) and (f). Values followed by $^*$ were extracted from local fits in the 5.5-7 keV band, with the continuum modelled as an absorbed powerlaw, with photon index and column density fixed to the best fit values in the broad band fit. See text for details.} 
% use packages: array
\begin{tabular}{lllll}
\hline
& \multicolumn{3}{c}{\textit{Chandra}} & XMM-\textit{Newton}\\
 & Mrk~463E & Mrk~463W & Extended & Mrk~463E+W \\ 
\hline
 &  &  &  &\\
N$_\mathrm{H,1}$ ($10^{22}$ cm$^{-2}$) & $0.09\pm0.07$ & $<0.15$ & $<0.11$ & $<0.03$ \\
N$_\mathrm{H,2}$ ($10^{22}$ cm$^{-2}$) & $71^{+18}_{-15}$ & $32^{+9}_{-6}$ & -- & $55^{+10}_{-7}$ \\ 
$\Gamma$ & $2.3^{+0.3}_{-0.2}$ & $2.2\pm0.2$ & $2.3\pm0.3$ & $2.04^{+0.09}_{-0.11}$ \\
F$_{6.4}$ & $6.2^{+4.4}_{-3.6}$ $^*$ & $<2.0$ & -- & $9.4^{+3.1}_{-2.7}$ \\
EW$_{6.4}$ (eV) & $210^{+150}_{-120}$ $^*$& $<260$ $^*$ & -- & $280\pm90$ \\
$\chi^2$/dof & 25/29 & 9/10 & 13/12 & 76/94 \\
  &  &  &  & \\
$F_{0.5-2}$ & $0.5\pm0.1$ & $0.11\pm0.04$ & $0.21\pm0.04$ & $1.1\pm0.1$\\
$F_{2-10}$ & $4.1\pm1.8$ & $1.8\pm0.8$ & -- & $6.9\pm2.0$\\
$L_{2-10}$ & $15\pm7$ & $3.8\pm1.7$ & -- & $23\pm7$\\
  &  &  & &  \\
\multicolumn{4}{c}{\textsc{Fluxes of other emission lines}}\\
  &  &  & &  \\
{O\,\textsc{vii}} K$\alpha$ (rif) (0.568 keV) & -- & -- & -- & $8.7^{+0.5}_{-0.7}$ \\
{O\,\textsc{viii}} K$\alpha$ (0.654 keV) & $3.5^{+5.5}_{-3.4}$ & -- & -- & $4.8^{+2.1}_{-3.0}$ \\
{O\,\textsc{vii}} RRC (0.739 keV) & -- & -- & -- & $6.0^{+1.7}_{-2.4}$ \\
{Fe\,\textsc{xvii}} 3d-2p (0.826 keV) & $3.5^{+2.3}_{-2.0}$ & -- & $3.1^{+1.2}_{-1.0}$ & $7.2^{+1.6}_{-1.9}$ \\
{Ne\,\textsc{ix}} K$\alpha$ (rif) (0.914 keV) & $3.1^{+1.7}_{-1.6}$ & -- & -- & $7.3^{+1.4}_{-1.6}$ \\
{Ne\,\textsc{x}} K$\alpha$ (1.022 keV) & $2.0\pm1.1$ & $0.9\pm0.6$ & $1.7\pm0.6$ & $5.7^{+1.1}_{-1.3}$ \\
{Ne\,\textsc{x}} K$\beta$ (1.211 keV) & -- & -- & -- & $1.3^{+0.8}_{-0.9}$ \\
{Mg\,\textsc{xi}} K$\alpha$ (rif) (1.342 keV) & -- & -- & -- & $1.4\pm0.7$ \\
{Si\,\textsc{xiii}} K$\alpha$ (rif) (1.853 keV) & $0.5\pm0.4$ & -- & -- & $0.7^{+0.6}_{-0.5}$ \\
{Ar\,\textsc{xvii}} K$\alpha$ (rif) (3.123 keV) & -- & -- & -- & $1.1^{+0.7}_{-0.6}$ \\
{Ar\,\textsc{xvii}} K$\beta$ (3.680 keV) & -- & -- & -- & $1.7\pm0.7$ \\
{Ca\,\textsc{xx}} K$\alpha$ (4.070 keV) & $1.2\pm0.6$ & -- & -- & -- \\
Unid. ($4.31\pm0.04$ keV) & -- & -- & -- & $1.6^{+0.9}_{-0.8}$ \\
{Fe\,\textsc{xxvi}} (6.96 keV)  & $1.6^{+1.3}_{-0.9}$ $^*$& -- & -- & $3.4\pm2.3$ \\
&  &  & &  \\
\hline
\end{tabular}
\end{table*}

The 2-10 keV absorption-corrected luminosities strongly indicate that both nuclei are AGN, being $1.5\times10^{43}$ and $3.8\times10^{42}$ erg s$^{-1}$, much larger than observed in Starburst galaxies and LINERs \citep[e.g. $<\mathrm{L_x^{SB}}>=8.5\times10^{39}$ and $<\mathrm{L_x^{L}}>=1.7\times10^{40}$ erg s$^{-1}$:][]{gonzalez06}. In addition, the presence of a strong iron line (EW$\simeq210$ eV) is clearly detected in the local fit of the brightest nucleus, while the upper limit found for the other source is largely consistent with what expected for an AGN (see Table \ref{xmmfit}).

The XMM-\textit{Newton} spectrum comprises the sum of the two nuclei and the extended emission. Indeed, the best fit is what expected from a Compton-thin Seyfert 2, with a strong Fe K$\alpha$ line at high energies and several emission lines from lighter metals in the soft X-rays. The column density of the absorber is intermediate between the two measured separately for the two nuclei with \textit{Chandra}. The 0.5-2 and 2-10 keV fluxes are consistent, within errors, with the sum of all the contributions measured with \textit{Chandra}. The neutral iron line flux is larger than the one observed for nucleus East, suggesting the intriguing possibility that the remaining flux comes from an otherwise undetected iron line flux in nucleus West. However, the first two fluxes are consistent within errors and so nothing conclusive can be said in this respect.

As found in the \textit{Chandra} data, the soft X-ray emission is dominated by a wealth of emission lines from ionised metals. The much better quality of the XMM-\textit{Newton} data allows us to detect the {O\,\textsc{vii}} RRC, a clear signature of photoionisation \citep[see e.g.][and references intherein]{gb07}. Finally, a strong {Fe\,\textsc{xxvi}} line is detected \citep[as already reported by][]{it04}, confirming the one observed in the \textit{Chandra} East nucleus (although, in the latter case, the centroid energy is marginally inconsistent with 6.96 keV and may be blended with {Fe\,\textsc{xxv}}). Emission lines from ionised iron are common in Seyfert 2s and may originate in a more ionised phase of the circumnuclear gas photoionised by the nuclear continuum \citep[see e.g.][]{bianchi05}.

\section{Discussion}

The detection in the \textit{Chandra} data of two luminous ($\mathrm{L}_\mathrm{2-10\,keV}=1.5\times10^{43}$ and $3.8\times10^{42}$ erg cm$^{-2}$ s$^{-1}$), unresolved nuclei in Mrk~463 strongly suggests that this galaxy hosts a binary AGN. While the East nucleus was already known to be a Seyfert 2 (and this is further confirmed by our \textit{Chandra} detection of a neutral iron line), this is the first unambiguous evidence in favour of the AGN nature of the West nucleus. These results make Mrk~463 the best \textit{bona-fide} binary AGN, only after NGC~6240, where strong neutral iron lines are detected in both nuclei. The projected separation between the two nuclei is $\simeq3.8$ kpc, again larger only than the one measured in NGC~6240. 

\begin{figure}
\begin{center}
\epsfig{file=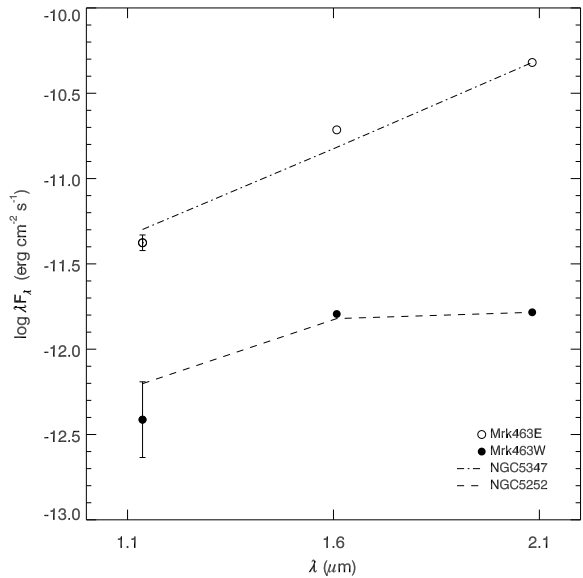, width=8cm}
\end{center}
\caption{\label{sed}The \textit{HST} NIR spectrum of the two nuclei in Mrk~463. In comparison, the \textit{HST} NIR spectrum of two Seyfert 2 galaxies are plotted, re-normalised at the 2.1 $\mu$m flux of the nuclei. See text for details.}
\end{figure}

Fig. \ref{sed} shows the NIR spectrum of the two nuclei, using the \textit{HST} data. In the same figure, we also plotted the \textit{HST} NIR spectrum of two typical Seyfert 2 galaxies taken from \citet{alherr03}, re-normalized to the 2.1 $\mu$m flux of the two nuclei. The spectrum of Mrk~463E is reasonably well fitted by that of Mrk~573, a Compton-thick source \citep{gua05b}, while Mrk~463W better fits that of a Compton-thin source, NGC~5252 \citep[N$_\mathrm{H}\simeq5-7\times10^{22}$ cm$^{-2}$:][]{risa02b}. Even if the column densities of these Seyfert 2s do not match those of the nuclei of Mrk~463 (and, in principle, they may vary), this comparison still supports that one of the nuclei is more obscured than the other, as found in the X-rays. More direct information on the dust absorption in Mrk~463E and its relation to the X-ray obscuration could come, in principle, from IR spectroscopy. Indeed, the optical depth of the 3.4 $\mu$m carbonaceous dust absorption feature allowed \citet{ima02} to estimate a dust extinction $\mathrm{A_V\simeq6-11}$. With the standard Galactic gas-to-dust ratio \citep[see e.g.][and references therein]{mai01}, this value would correspond to $\mathrm{N_H\simeq2\times10^{22}}$ cm$^2$, which is a factor of $\simeq35$ lower than our X-ray measure. However, such an inconsistency between dust extinction and X-ray absorption is very common in AGN \citep{mai01} and so a direct comparison between the two is difficult to interpret.

Adopting the \textit{reddening-corrected} fluxes reported by \citet{mb93}, we get [O\,\textsc{iii}] luminosities of $5.4\times10^{42}$ and $3.2\times10^{41}$ erg s$^{-1}$ for nucleus E and W, respectively. Therefore, with respect to the \textit{unabsorbed} 2-10 keV X-ray luminosities derived in our \textit{Chandra} analysis, Mrk~463W has $\log{\mathrm{X}/{\mathrm{[O}\,\textsc{iii}\mathrm{]}}}=1.07$, which is in agreement with unobscured or Compton-thin Seyfert galaxies \citep[e.g.][]{panessa06}. On the other hand, Mrk~463E has  $\log{\mathrm{X}/{\mathrm{[O}\,\textsc{iii}\mathrm{]}}}=0.44$, making it X-ray under-luminous, unless it is Compton-thick, but this is quite unlikely given the low EW of the iron line. Indeed, ULIRG are often found to be X-ray under-luminous, but it is unclear whether it is an intrinsic property of AGN in ULIRG or it is due to the larger amount of dust available in these systems \citep[e.g.][]{brandt97,it04}. The interesting point here is that only one of the two active nuclei of this ULIRG is X-ray under-luminous and it is the one which presents an excess with respect to the other in IR ($\simeq30$) and [O\,\textsc{iii}] ($\simeq17$) emission, in comparison with the ratio of their intrinsic X-ray luminosities, which is only 4. It is important to note here that almost all of the IR emission of Mrk~463E comes from the AGN, since any compact nuclear starbursts in this source are energetically insignificant, as found on the basis of the analysis of its IR spectrum \citep{ima02,armus04}.

However, it must be noted that the [O\,\textsc{iii}], IR and X-ray observations are not simultaneous and care must be taken when comparing them, without taking into account the source's intrinsic variability. With respect to the 1990 IR observations taken by \citet{mazz91}, Mrk~463E has dimmed by 0.55 magnitudes and Mrk~463W by 0.13 magnitudes during the \textit{HST} observation, taken 7 years later. While we do not have any evidence that any of the two nuclei varied in the X-rays (we have only the \textit{Chandra} observation analysed in this paper where the nuclei are resolved, and their combined X-ray flux is consistent with the one measured by \textit{ASCA} and XMM-\textit{Newton}), it is clearly possible that much (if not all) of the X-ray under-luminosity of Mrk~463E may be due to the comparison of two different states of the source.

On the basis of 16 optical binary candidates, \citet{mort99} estimated that the typical separation where the galactic nuclei become active during a merging process (the so-called `activation radius') lies between 50 and 100 kpc. Their model also predicts that, when the merger becomes more stable, at distances around 10 kpc, the nuclei should `turn-off' again, due to the end of the inflow of gas supplied by the merger in the previous phase. However, it is interesting to note that, with the possible exception of AM1211-465 \citep{jb07}, all the binary AGN found in the X-rays have instead small apparent separation distances, less than $\simeq10$ kpc.

Moreover, as already noted by \citet{gua05c}, all the AGN pairs observed so far are found to be heavily obscured in the X-rays, if not Compton-thick, and the two nuclei in Mrk~463 confirm this trend. Even if based on small numbers, this result is in agreement with the idea that merging of galaxies may trigger the presence of a large amount of gas in the nuclear environment. Since obscuration may prevent a correct classification in other bands, hard X-ray observations are indeed the most effective probes to find these elusive systems, which are expected to be abundant if our understanding of the formation and evolution of galaxies and AGN is correct.

\section*{Acknowledgements}

SB, EP and GM acknowledge financial support from ASI (grant I/088/06/0). We thank the anonymous referee for useful suggestions.

\bibliographystyle{mn}
\bibliography{sbs}

\begin{thebibliography}{39}
\expandafter\ifx\csname natexlab\endcsname\relax\def\natexlab#1{#1}\fi

\bibitem[{{Alonso-Herrero} {et~al.}(2003){Alonso-Herrero}, {Quillen}, {Rieke},
  {Ivanov}, \& {Efstathiou}}]{alherr03}
{Alonso-Herrero} A., {Quillen} A.~C., {Rieke} G.~H., {Ivanov} V.~D.,
  {Efstathiou} A., 2003, \aj, 126, 81

\bibitem[{{Armus} {et~al.}(2004){Armus}, {Charmandaris}, {Spoon}, {Houck},
  {Soifer}, {Brandl}, {Appleton}, {Teplitz}, {Higdon}, {Weedman}, {Devost},
  {Morris}, {Uchida}, {van Cleve}, {Barry}, {Sloan}, {Grillmair}, {Burgdorf},
  {Fajardo-Acosta}, {Ingalls}, {Higdon}, {Hao}, {Bernard-Salas}, {Herter},
  {Troeltzsch}, {Unruh}, \& {Winghart}}]{armus04}
{Armus} L., et al., 2004, \apjs, 154, 178

\bibitem[{{Arnaud}(1996)}]{xspec}
{Arnaud} K.~A., 1996, in ASP Conf. Ser. 101: Astronomical Data Analysis
  Software and Systems V, p.~17

\bibitem[{{Awaki} \& {Koyama}(1993)}]{awaki93}
{Awaki} H., {Koyama} K., 1993, Advances in Space Research, 13, 221

\bibitem[{{Ballo} {et~al.}(2004){Ballo}, {Braito}, {Della Ceca}, {Maraschi},
  {Tavecchio}, \& {Dadina}}]{ballo04}
{Ballo} L., {Braito} V., {Della Ceca} R., {Maraschi} L., {Tavecchio} F.,
  {Dadina} M., 2004, \apj, 600, 634

\bibitem[{{Barker} {et~al.}(2007){Barker}, {Barker}, {Barker}, \&
  {Barker}}]{barker07}
{Barker} E., {Barker} E., {Barker} E., {Barker} E., 2007, {"NICMOS Instrument
  Handbook", Version 10.0, (Baltimore: STScI)}

\bibitem[{{Bianchi} {et~al.}(2006){Bianchi}, {Guainazzi}, \&
  {Chiaberge}}]{bianchi06}
{Bianchi} S., {Guainazzi} M., {Chiaberge} M., 2006, \aap, 448, 499

\bibitem[{{Bianchi} {et~al.}(2005){Bianchi}, {Matt}, {Nicastro}, {Porquet}, \&
  {Dubau}}]{bianchi05}
{Bianchi} S., {Matt} G., {Nicastro} F., {Porquet} D., {Dubau} J., 2005, \mnras,
  357, 599

\bibitem[{{Brandt} {et~al.}(1997){Brandt}, {Fabian}, {Takahashi}, {Fujimoto},
  {Yamashita}, {Inoue}, \& {Ogasaka}}]{brandt97}
{Brandt} W.~N., {Fabian} A.~C., {Takahashi} K., {Fujimoto} R., {Yamashita} A.,
  {Inoue} H., {Ogasaka} Y., 1997, \mnras, 290, 617

\bibitem[{{Cash}(1976)}]{cash76}
{Cash} W., 1976, \aap, 52, 307

\bibitem[{{Dickey} \& {Lockman}(1990)}]{dl90}
{Dickey} J.~M., {Lockman} F.~J., 1990, \araa, 28, 215

\bibitem[{{Evans} {et~al.}(2007){Evans}, {Fong}, {Hardcastle}, {Kraft}, {Lee},
  {Worrall}, {Birkinshaw}, {Croston}, \& {Muxlow}}]{evans08}
{Evans} D.~A., et al., 2007, ArXiv e-prints, 712

\bibitem[{{Fruscione} {et~al.}(2006){Fruscione}, {McDowell}, {Allen},
  {Brickhouse}, {Burke}, {Davis}, {Durham}, {Elvis}, {Galle}, {Harris},
  {Huenemoerder}, {Houck}, {Ishibashi}, {Karovska}, {Nicastro}, {Noble},
  {Nowak}, {Primini}, {Siemiginowska}, {Smith}, \& {Wise}}]{ciao}
{Fruscione} A., et al.., 2006, in Presented at the Society of
  Photo-Optical Instrumentation Engineers (SPIE) Conference, Vol. 6270,
  Observatory Operations: Strategies, Processes, and Systems. Edited by Silva,
  David R.; Doxsey, Rodger E.. Proceedings of the SPIE, Volume 6270, pp. 62701V
  (2006).

\bibitem[{{Gabriel} {et~al.}(2004){Gabriel}, {Denby}, {Fyfe}, {Hoar}, {Ibarra},
  {Ojero}, {Osborne}, {Saxton}, {Lammers}, \& {Vacanti}}]{sas610}
{Gabriel} C., et al., 2004, in ASP Conf.
  Ser. 314: Astronomical Data Analysis Software and Systems (ADASS) XIII, p.
  759

\bibitem[{{Garmire} {et~al.}(2003){Garmire}, {Bautz}, {Ford}, {Nousek}, \&
  {Ricker}}]{acis}
{Garmire} G.~P., {Bautz} M.~W., {Ford} P.~G., {Nousek} J.~A., {Ricker} G.~R.,
  2003, in X-Ray and Gamma-Ray Telescopes and Instruments for Astronomy. Edited
  by Joachim E. Truemper, Harvey D. Tananbaum. Proceedings of the SPIE, Volume
  4851, p. 28-44

\bibitem[{{Gonz{\'a}lez-Mart{\'{\i}}n}
  {et~al.}(2006){Gonz{\'a}lez-Mart{\'{\i}}n}, {Masegosa}, {M{\'a}rquez},
  {Guerrero}, \& {Dultzin-Hacyan}}]{gonzalez06}
{Gonz{\'a}lez-Mart{\'{\i}}n} O., {Masegosa} J., {M{\'a}rquez} I., {Guerrero}
  M.~A., {Dultzin-Hacyan} D., 2006, \aap, 460, 45

\bibitem[{{Guainazzi} \& {Bianchi}(2007)}]{gb07}
{Guainazzi} M., {Bianchi} S., 2007, \mnras, 374, 1290

\bibitem[{{Guainazzi} {et~al.}(2005{\natexlab{a}}){Guainazzi}, {Matt}, \&
  {Perola}}]{gua05b}
{Guainazzi} M., {Matt} G., {Perola} G.~C., 2005{\natexlab{a}}, \aap, 444, 119

\bibitem[{{Guainazzi} {et~al.}(2005{\natexlab{b}}){Guainazzi}, {Piconcelli},
  {Jim{\'e}nez-Bail{\'o}n}, \& {Matt}}]{gua05c}
{Guainazzi} M., {Piconcelli} E., {Jim{\'e}nez-Bail{\'o}n} E., {Matt} G.,
  2005{\natexlab{b}}, \aap, 429, L9

\bibitem[{{Haiman} \& {Quataert}(2004)}]{hq04}
{Haiman} Z., {Quataert} E., 2004, in Astrophysics and Space Science Library,
  Vol. 308, Supermassive Black Holes in the Distant Universe, {Barger} A.~J.,
  ed., pp. 147--+

\bibitem[{{Hutchings} \& {Neff}(1989)}]{hn89}
{Hutchings} J.~B., {Neff} S.~G., 1989, \aj, 97, 1306

\bibitem[{{Imanishi}(2002)}]{ima02}
{Imanishi} M., 2002, \apj, 569, 44

\bibitem[{{Imanishi} \& {Terashima}(2004)}]{it04}
{Imanishi} M., {Terashima} Y., 2004, \aj, 127, 758

\bibitem[{{Jim{\'e}nez-Bail{\'o}n} {et~al.}(2007){Jim{\'e}nez-Bail{\'o}n},
  {Loiseau}, {Guainazzi}, {Matt}, {Rosa-Gonz{\'a}lez}, {Piconcelli}, \&
  {Santos-Lle{\'o}}}]{jb07}
{Jim{\'e}nez-Bail{\'o}n} E., {Loiseau} N., {Guainazzi} M., {Matt} G.,
  {Rosa-Gonz{\'a}lez} D., {Piconcelli} E., {Santos-Lle{\'o}} M., 2007, \aap,
  469, 881

\bibitem[{{Komossa}(2003)}]{kom03}
{Komossa} S., 2003, in American Institute of Physics Conference Series, Vol.
  686, The Astrophysics of Gravitational Wave Sources, {Centrella} J.~M., ed.,
  pp. 161--174

\bibitem[{{Komossa} {et~al.}(2003){Komossa}, {Burwitz}, {Hasinger}, {Predehl},
  {Kaastra}, \& {Ikebe}}]{kom03b}
{Komossa} S., {Burwitz} V., {Hasinger} G., {Predehl} P., {Kaastra} J.~S.,
  {Ikebe} Y., 2003, \apjl, 582, L15

\bibitem[{{Landi} \& {Bassani}(2001)}]{lb01}
{Landi} R., {Bassani} L., 2001, \aap, 379, 855

\bibitem[{{Maiolino} {et~al.}(2001){Maiolino}, {Marconi}, {Salvati},
  {Risaliti}, {Severgnini}, {Oliva}, {La Franca}, \& {Vanzi}}]{mai01}
{Maiolino} R., {Marconi} A., {Salvati} M., {Risaliti} G., {Severgnini} P.,
  {Oliva} E., {La Franca} F., {Vanzi} L., 2001, \aap, 365, 28

\bibitem[{{Mazzarella} \& {Boroson}(1993)}]{mb93}
{Mazzarella} J.~M., {Boroson} T.~A., 1993, \apjs, 85, 27

\bibitem[{{Mazzarella} {et~al.}(1991){Mazzarella}, {Soifer}, {Graham},
  {Neugebauer}, {Matthews}, \& {Gaume}}]{mazz91}
{Mazzarella} J.~M., {Soifer} B.~T., {Graham} J.~R., {Neugebauer} G., {Matthews}
  K., {Gaume} R.~A., 1991, \aj, 102, 1241

\bibitem[{{Miller} \& {Goodrich}(1990)}]{mg90}
{Miller} J.~S., {Goodrich} R.~W., 1990, \apj, 355, 456

\bibitem[{{Milosavljevi{\'c}} \& {Merritt}(2001)}]{mm01}
{Milosavljevi{\'c}} M., {Merritt} D., 2001, \apj, 563, 34

\bibitem[{{Mortlock} {et~al.}(1999){Mortlock}, {Webster}, \&
  {Francis}}]{mort99}
{Mortlock} D.~J., {Webster} R.~L., {Francis} P.~J., 1999, \mnras, 309, 836

\bibitem[{{Panessa} {et~al.}(2006){Panessa}, {Bassani}, {Cappi}, {Dadina},
  {Barcons}, {Carrera}, {Ho}, \& {Iwasawa}}]{panessa06}
{Panessa} F., {Bassani} L., {Cappi} M., {Dadina} M., {Barcons} X., {Carrera}
  F.~J., {Ho} L.~C., {Iwasawa} K., 2006, \aap, 455, 173

\bibitem[{{Piconcelli} {et~al.}(2004){Piconcelli}, {Jimenez-Bail{\' o}n},
  {Guainazzi}, {Schartel}, {Rodr{\'{\i}}guez-Pascual}, \& {Santos-Lle{\'
  o}}}]{pico04}
{Piconcelli} E., {Jimenez-Bail{\' o}n} E., {Guainazzi} M., {Schartel} N.,
  {Rodr{\'{\i}}guez-Pascual} P.~M., {Santos-Lle{\' o}} M., 2004, \mnras, 351,
  161

\bibitem[{{Polletta} {et~al.}(1996){Polletta}, {Bassani}, {Malaguti},
  {Palumbo}, \& {Caroli}}]{polletta96}
{Polletta} M., {Bassani} L., {Malaguti} G., {Palumbo} G.~G.~C., {Caroli} E.,
  1996, \apjs, 106, 399

\bibitem[{{Risaliti} {et~al.}(2002){Risaliti}, {Elvis}, \&
  {Nicastro}}]{risa02b}
{Risaliti} G., {Elvis} M., {Nicastro} F., 2002, \apj, 571, 234

\bibitem[{{Sanders} \& {Mirabel}(1996)}]{sm96}
{Sanders} D.~B., {Mirabel} I.~F., 1996, \araa, 34, 749

\bibitem[{{Shuder} \& {Osterbrock}(1981)}]{so81}
{Shuder} J.~M., {Osterbrock} D.~E., 1981, \apj, 250, 55

\bibitem[{{Str{\"u}der} {et~al.}(2001){Str{\"u}der}, {Briel}, {Dennerl},
  {Hartmann}, {Kendziorra}, {Meidinger}, {Pfeffermann}, {Reppin}, {Aschenbach},
  {Bornemann}, {Br{\" a}uninger}, {Burkert}, {Elender}, {Freyberg}, {Haberl},
  {Hartner}, {Heuschmann}, {Hippmann}, {Kastelic}, {Kemmer}, {Kettenring},
  {Kink}, {Krause}, {M{\" u}ller}, {Oppitz}, {Pietsch}, {Popp}, {Predehl},
  {Read}, {Stephan}, {St{\" o}tter}, {Tr{\" u}mper}, {Holl}, {Kemmer},
  {Soltau}, {St{\" o}tter}, {Weber}, {Weichert}, {von Zanthier},
  {Carathanassis}, {Lutz}, {Richter}, {Solc}, {B{\" o}ttcher}, {Kuster},
  {Staubert}, {Abbey}, {Holland}, {Turner}, {Balasini}, {Bignami}, {La
  Palombara}, {Villa}, {Buttler}, {Gianini}, {Lain{\' e}}, {Lumb}, \&
  {Dhez}}]{struder01}
{Str{\"u}der} L., et al., 2001, \aap, 365, L18

\bibitem[{{Turner} {et~al.}(2001){Turner}, {Abbey}, {Arnaud}, {Balasini},
  {Barbera}, {Belsole}, {Bennie}, {Bernard}, {Bignami}, {Boer}, {Briel},
  {Butler}, {Cara}, {Chabaud}, {Cole}, {Collura}, {Conte}, {Cros}, {Denby},
  {Dhez}, {Di Coco}, {Dowson}, {Ferrando}, {Ghizzardi}, {Gianotti}, {Goodall},
  {Gretton}, {Griffiths}, {Hainaut}, {Hochedez}, {Holland}, {Jourdain},
  {Kendziorra}, {Lagostina}, {Laine}, {La Palombara}, {Lortholary}, {Lumb},
  {Marty}, {Molendi}, {Pigot}, {Poindron}, {Pounds}, {Reeves}, {Reppin},
  {Rothenflug}, {Salvetat}, {Sauvageot}, {Schmitt}, {Sembay}, {Short},
  {Spragg}, {Stephen}, {Str{\" u}der}, {Tiengo}, {Trifoglio}, {Tr{\" u}mper},
  {Vercellone}, {Vigroux}, {Villa}, {Ward}, {Whitehead}, \& {Zonca}}]{turner01}
{Turner} M.~J.~L., et al., 2001, \aap, 365, L27

\bibitem[{{Ueno} {et~al.}(1996){Ueno}, {Koyama}, {Awaki}, {Hayashi}, \&
  {Blanco}}]{ueno96}
{Ueno} S., {Koyama} K., {Awaki} H., {Hayashi} I., {Blanco} P.~R., 1996, \pasj,
  48, 389

\end{thebibliography}

\label{lastpage}

\end{document}